\documentclass{article}
\usepackage{graphicx}

\def\fracd#1#2{{\displaystyle\frac{#1}{#2}}}

\begin{document}
\noindent\underline{Astronomy Reports, 2013, Vol. 57, No. 2, pp. 128--133. DOI: 10.1134/S1063772913020078}

\vskip 5em
\centerline{\LARGE\bf Analysis of Determinations of the Distance \strut}
\centerline{\LARGE\bf between the Sun and the Galactic Center \strut}
\bigskip
\centerline{\Large\bf Z. M. Malkin}
\bigskip
\centerline{\large Pulkovo Observatory, St. Petersburg, Russia}
\centerline{\large St. Petersburg State University, St. Petersburg, Russia}
\centerline{\large e-mail: malkin@gao.spb.ru}

\begin{abstract}
The paper investigates the question of whether or not determinations of the distance between
the Sun and the Galactic center $R_0$ are affected by the so-called "bandwagon effect", leading to selection
effects in published data that tend to be close to expected values, as was suggested by some authors. It
is difficult to estimate numerically a systematic uncertainty in $R_0$ due to the bandwagon effect; however, it
is highly probable that, even if widely accepted values differ appreciably from the true value, the published
results should eventually approach the true value despite the bandwagon effect. This should be manifest
as a trend in the published $R_0$ data: if this trend is statistically significant, the presence of the bandwagon
effect can be suspected in the data. Fifty two determinations of $R_0$ published over the last 20 years were
analyzed. These data reveal no statistically significant trend, suggesting they are unlikely to involve any
systematic uncertainty due to the bandwagon effect. At the same time, the published data show a gradual
and statistically significant decrease in the uncertainties in the $R_0$ determinations with time.
\end{abstract}

\section{Introduction}

Accurate knowledge of the distance between the
Sun and the Galactic center $R_0$ is very important
for many problems associated with the structure and
evolution of the Universe, as well as for astrometric
applications. For example, the need to determine the
effect of Galactic aberration on the proper motions of
radio sources more precisely motivated our study of
the accuracy of $R_0$ determinations [1].

Like any quantity derived from observations, determinations
of $R_0$ have both random and systematic
uncertainties. The systematic uncertainties are most
dangerous in terms of using the corresponding results
in various applications. Such systematic uncertainties
could be due to instrumental and methodical uncertainties,
or uncertainties in a theory used to interpret
the measured values. Alongside these, subjective
uncertainties that are not associated with the accuracy
of the observations or data processing can arise.
One example is the "bandwagon effect". This refers to
the possibility that only measured results corresponding
fairly well to the expected values are published;
i.e., these are results that differ from commonly accepted
expectations (or values published earlier) by a
``reasonable'' (usually small) amount. However, the
fact that the overwhelming majority of published data
were honestly derived from observations suggests that, even if they are filtered by publication selection,
published values should gradually approach the true
value. Therefore, the presence of a trend in the published
determinations of some quantity may indicate a
role played by the bandwagon effect (naturally, taking
into account possible real variations of this quantity
with time).

In our case, a role of the bandwagon effect has
been suspected in a number of reviews devoted to
$R_0$ determinations (e.g., [2-4]), which can be significantly
subject to this systematic uncertainty. These
suspicions were based on claims of systematic variation
of the published $R_0$ values with time. However,
the trends obtained in [2-4] do not agree with each
other (this is discussed in Section 3 in more detail).
Moreover, many results refer to out-of-date publications
that are almost irrelevant now. The current
study is aimed at elucidating whether or not published
values of $R_0$ obtained over the past 20 years display
appreciable time-dependent variations.

\section{Determinations of $R_0$ used in the analysis}
\label{sect_input}

We used $R_0$ determinations published between
1992 and 2011. Those interested in earlier data can
refer to the reviews [2,3,5,6]. We used all published
data except those refined later. Thus, we did not use
the result of [7], revised in [8], the result of [9], revised
in [10], and the result of [11], revised in [12] and later in [13].

In cases when the estimates of both random ($\varepsilon_{stat}$) and systematic ($\varepsilon_{syst}$) uncertainties are available, we
calculated the total uncertainty as the square root of the sum of their squares, $\varepsilon = \sqrt{\varepsilon_{stat}^2 + \varepsilon_{syst}^2}$.
The
corresponding data are presented in Table~\ref{tab:stat-syst}.

\begin{table}[ht!]
\centering
\caption{Values of $R_0$ for which random and systematic uncertainties are available}
\label{tab:stat-syst}
\begin{tabular}{cc}
\hline
\multicolumn{1}{c}{Reference} & $R_0$, kpc \\
\hline
$[14]$ & $R_0 = 7.52 \pm 0.10\,|_{stat} \pm 0.35\,|_{syst}$ \\
$[15]$ & $R_0 = 7.94 \pm 0.37\,|_{stat} \pm 0.26\,|_{syst}$ \\
$[16]$ & $R_0 = 8.07 \pm 0.32\,|_{stat} \pm 0.13\,|_{syst}$ \\
$[13]$ & $R_0 = 8.33 \pm 0.17\,|_{stat} \pm 0.31\,|_{syst}$ \\
$[17]$ & $R_0 = 8.28 \pm 0.15\,|_{stat} \pm 0.29\,|_{syst}$ \\
$[18]$ & $R_0 = 8.24 \pm 0.08\,|_{stat} \pm 0.42\,|_{syst}$ \\
$[19]$ & $R_0 = 8.3\phantom{3} \pm 0.46\,|_{stat} \pm 1.0\phantom{3}\,|_{syst}$ \\
\hline
\end{tabular}
\end{table}

When two versions of the $R_0$ determinations are available, we averaged these values. These results are
given in Table~\ref{tab:averaged}.

\begin{table}[ht!]
\centering
\caption{Mean values of $R_0$}
\label{tab:averaged}
\begin{tabular}{ccc}
\hline
\multicolumn{1}{c}{Reference} & \multicolumn{2}{c}{$R_0$, kpc} \\
\hline
$[20]$ & $8.7  \pm 0.7$  & $8.9  \pm 0.7$ \\
$[21]$ & $7.6  \pm 0.4$  & $8.3  \pm 0.5$ \\
$[22]$ & $7.9  \pm 0.85$ & $8.2  \pm 0.9$ \\
$[23]$ & $8.6  \pm 0.7$  & $8.8  \pm 0.4$ \\
$[24]$ & $7.95 \pm 0.62$ & $8.25 \pm 0.79$ \\
$[25]$ & $7.96 \pm 0.63$ & $8.36 \pm 0.37$ \\
$[26]$ & $7.7  \pm 0.7$  & $7.8  \pm 0.6$ \\
\hline
\end{tabular}
\end{table}

Asymmetric confidence intervals, rather than rms uncertainties, are provided as estimates of the accuracy
in some studies [25,27,28]. In these cases,
the uncertainties are taken to be the mean values of
the lower and upper bounds of the intervals. Since
these bounds are close to each other in all cases, this
substitution does not appreciably affect our results.

The final list of all the $R_0$ values used is presented
in Table~\ref{tab:allr0}.

\begin{table}[ht!]
\centering
\caption{Results of $R_0$ determinations}
\label{tab:allr0}
\begin{tabular}{llc|cllc}
\hline
$R_0$, kpc & STD & Reference && $R_0$, kpc & STD & Reference \\
\hline
7.9  & 0.8   & $[29]$ &&  8.05 & 0.6   & $[22]$ \\
8.1  & 1.1   & $[30]$ &&  8.3  & 0.3   & $[31]$ \\
7.6  & 0.6   & $[32]$ &&  7.7  & 0.15  & $[33]$ \\
7.6  & 0.4   & $[34]$ &&  8.01 & 0.44  & $[35]$ \\
8.09 & 0.3   & $[36]$ &&  8.7  & 0.6   & $[23]$ \\
7.5  & 1.0   & $[37]$ &&  7.2  & 0.3   & $[38]$ \\
7.0  & 0.5   & $[39]$ &&  7.52 & 0.36  & $[14]$ \\
8.8  & 0.5   & $[20]$ &&  8.1  & 0.7   & $[24]$ \\
7.1  & 0.5   & $[40]$ &&  7.4  & 0.3   & $[41]$ \\
8.3  & 1.0   & $[42]$ &&  7.94 & 0.45  & $[15]$ \\
8.21 & 0.98  & $[43]$ &&  8.07 & 0.35  & $[16]$ \\
7.95 & 0.4   & $[21]$ &&  8.16 & 0.5   & $[25]$ \\
7.55 & 0.7   & $[44]$ &&  8.33 & 0.35  & $[13]$ \\
8.1  & 0.4   & $[45]$ &&  8.7  & 0.5   & $[28]$ \\
8.5  & 0.5   & $[46]$ &&  7.58 & 0.40  & $[47]$ \\
7.66 & 0.54  & $[48]$ &&  7.2  & 0.3   & $[49]$ \\
8.1  & 0.15  & $[50]$ &&  8.4  & 0.6   & $[51]$ \\
7.1  & 0.4   & $[52]$ &&  7.75 & 0.5   & $[26]$ \\
8.51 & 0.29  & $[53]$ &&  7.9  & 0.75  & $[27]$ \\
8.2  & 0.21  & $[54]$ &&  8.24 & 0.43  & $[18]$ \\
8.6  & 1.0   & $[6] $ &&  8.28 & 0.33  & $[17]$ \\
7.4  & 0.3   & $[8] $ &&  7.7  & 0.4   & $[55]$ \\
7.9  & 0.3   & $[56]$ &&  8.1  & 0.6   & $[57]$ \\
8.67 & 0.4   & $[10]$ &&  8.3  & 1.1   & $[19]$ \\
8.2  & 0.7   & $[58]$ &&  7.80 & 0.26  & $[59]$ \\
8.24 & 0.42  & $[60]$ &&  8.3  & 0.23  & $[61]$ \\
\hline
\end{tabular}
\end{table}

\section{Data analysis}
\label{sect:analys}

The existence of a trend in the published $R_0$ values
has been suggested in various earlier studies
[2-4,6,62]. A systematic time-dependent decrease
in published $R_0$ values during 1970-1990 was found
in [2,3,6,62]. Although these data are mainly of
historical interest now, this first suggestion of the
presence of a bandwagon effect in $R_0$ determinations
seems to have been based on a time-dependent drift
in the data [62]. At the same time, the data for
1990--1998 do not show any significant trend [6],
while the data of [3] show a slight positive trend,
i.e., a small increase in $R_0$ values with time during
1990--2003. A large positive trend was found in [4]
using results obtained in 1992--2010.

These results are somewhat contradictory, although
we must bear in mind that they sometimes
refer to different observing intervals. The conclusions
of the papers cited above are based on various samples
of published results, so that they could be distorted by
selection effects in the data used.

\begin{figure}[ht!]
\centering
\resizebox{0.77\hsize}{!}{\includegraphics[clip]{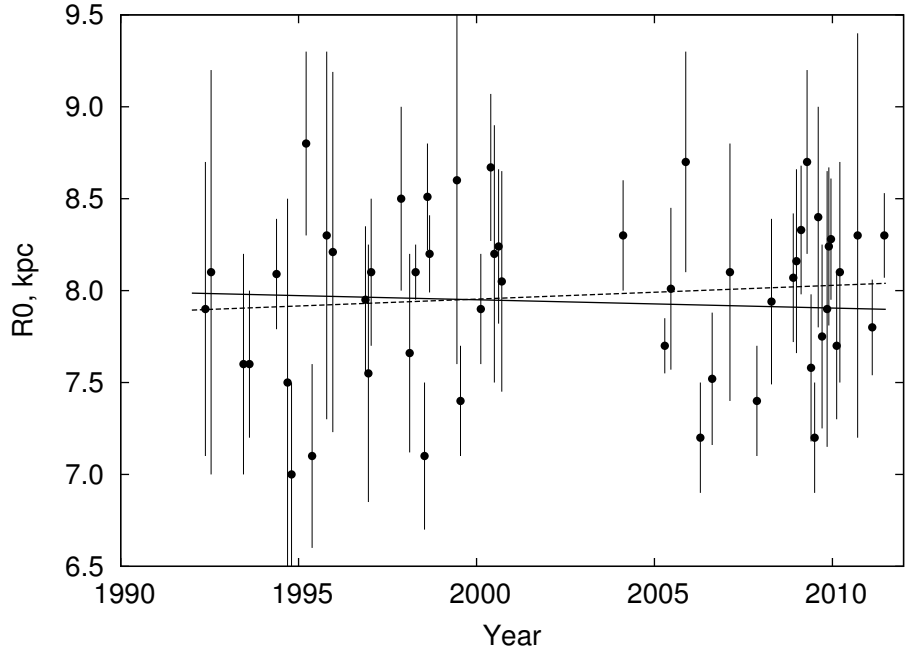}}
\caption{Values of $R_0$ used in this study. The solid and dashed lines correspond to the weighted and unweighted versions of the
calculated trends, respectively.}
\label{fig:data_R0}
\end{figure}

To resolve this contradiction, we analyzed all
available determinations of $R_0$ indicated in the previous
Section (see Fig.~\ref{fig:data_R0}). The epoch for a given $R_0$
value, taken to refer to the date of publication, is
plotted along the horizontal axis in Fig.~\ref{fig:data_R0}. This
epoch was determined by the month of publication
in the corresponding journal issue number, or taken
from the ADS bibliographic database (usually, for
conference proceedings). If it was not possible to
determine the month of publication, we adopted the
middle of the year as the date of publication. When the
epochs for two results coincided, we shifted them both
by 0.02--0.03 yr in opposite directions to separate the
corresponding points in the graph.

We calculated two versions of a linear trend using
the given data. In the first version, the weights
for the original data were taken to be equal to the
inverse squares of their uncertainties. The resulting
trend is $-0.0045 \pm 0.0103$ kpc/yr. We did not take
the weights into consideration in the second version,
resulting in the trend $+0.0075 \pm 0.0100$ kpc/yr.

A simple, but effective criterion can be used to
detect trends or other low-frequency variations in the
data, namely Abbe's criterion. This criterion is aimed
at testing the hypothesis that all the mathematical
expectancies of the analyzed measurements xi are
equal. To test this hypothesis, we calculated the
Abbe statistic, which is essentially the ratio of the
Allan variance $AV$ (whose applications to astronomical
studies are described in detail in [63,64]) to the
dispersion of the data $D$:
\begin{equation}
\begin{array}{rcl}
q &=& \fracd{AV}{D}\,, \\[1em]
AV &=& \fracd{\sum_{i=1}^{n-1}(y_i-y_{i+1})^2}{2(n-1)}\,, \\[1em]
D &=& \fracd{\sum_{i=1}^n {(x_i - \bar{x})^2}}{n-1} \,,
\end{array}
\end{equation}
where $\bar{x}$ is the mean of all the measurements $x_i$.

If there are appreciable low-frequency variations
in the data, including trends, $D$ will be appreciably
greater than the Allan variance. Thus, if $q$ is lower
than the corresponding critical point of the Abbe
distribution, the hypothesis that there is no trend
must be rejected, and we conclude that there are
statistically significant, systematic variations in the
data. Our value is much greater than the 1\% quantile for the Abbe distribution, which is equal to 0.69

Our conclusion based on all the calculations suggests
that there are no statistically significant trends
in the $R_0$ determinations for the last 20 years.

\begin{figure}[ht!]
\centering
\resizebox{0.75\hsize}{!}{\includegraphics[clip]{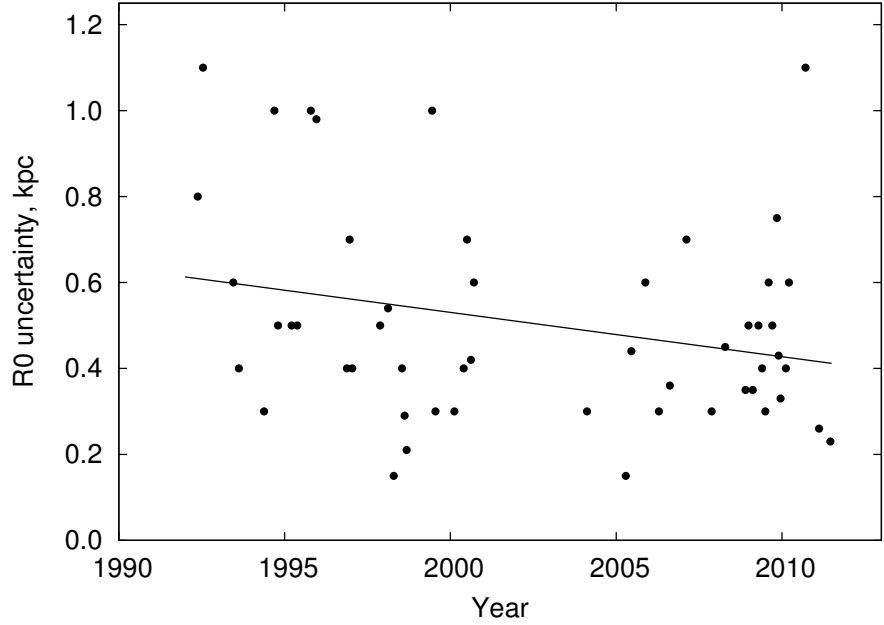}}
\caption{Uncertainties of the $R_0$ determinations, kpc.}
\label{fig:data_R0_err}
\end{figure}

It is also interesting to consider how the accuracy
of the $R_0$ determinations changes with time. This
analysis was carried out using the data shown in
Fig.~\ref{fig:data_R0_err}, yielding the trend $-0.0103 \pm 0.0053$~kpc/yr.
Thus, there is a statistically significant decrease in the
uncertainty of the $R_0$ determinations with time.

This conclusion seems all the more interesting
because researchers have recently begun to pay more
attention to estimate of the uncertainties of their results,
including analysis of both statistical (formal)
and systematic uncertainties. On the one hand, the
uncertainties in $R_0$ should decrease with time as the
observational data are accumulated and the observational
and data-processing techniques are improved;
on the other hand, these uncertainties could increase
as new, more correct methods are used to estimate the
accuracy. The former tendency has apparently been
stronger in recent years.

\section{Conclusions}

Our analysis of 52 $R_0$ determinations published
in 1992--2011 has shown that these data do not
display a statistically significant trend, i.e., a systematic
increase or decrease in the $R_0$ values with
time. Therefore, we do not confirm the conclusions
of Foster and Cooper [4], who found a large positive
trend, i.e., a systematic increase in $R_0$ values, over
the last 20 years. At the same time, we confirm the
results of [3,6] indicating a slight trend, although on
a shorter time interval.

\begin{figure}[ht!]
\centering
\resizebox{0.75\hsize}{!}{\includegraphics[clip]{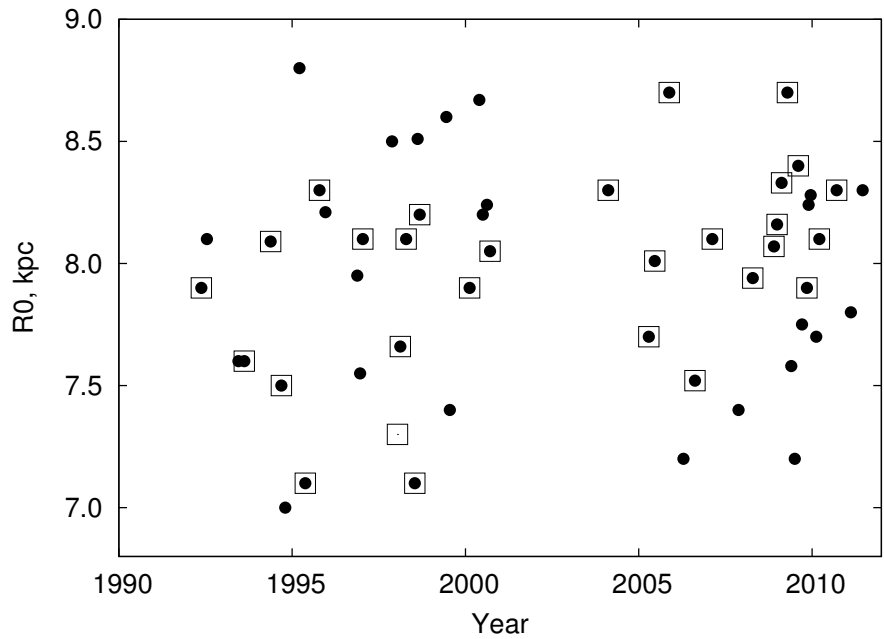}}
\caption{Sets of $R_0$ values used here and in [4] (the latter ones are denoted by squares), kpc.}
\label{fig:data_R0_f-c}
\end{figure}

The obvious origin of the discrepancy between our
results and those of [4] is selection effects associated
with the original data used in the analysis. This is
clearly visible in Fig.~\ref{fig:data_R0_f-c}, where our data and the data
of [4] are shown. For some reason, the latter data
did not include several $R_0$ values above the mean
during the first half of the interval, as well as several
values below the mean value during the second half
of the interval, yielding an appreciable increase in the
published $R_0$ values [4].

In our opinion, our main conclusion that there is
no statistically significant trend in the published $R_0$
values over the last 20 years suggests that these data
do not contain a statistically significant bandwagon
effect.

At the same time, we have discovered an appreciable
decrease in the uncertainties of the $R_0$ determinations
with time.

\clearpage
\section*{References}

\noindent
\leftskip=\parindent
\parindent=-\leftskip

1. Z. M. Malkin, Astron. Rep. 55, 810 (2011).

2. M. J. Reid, Ann. Rev. Astron. Astrophys. 31, 345
(1993).

3. I. Nikiforov, in Order and Chaos in Stellar and
Planetary Systems, Ed. by G. G. Byrd, K. V. Kholshevnikov,
A. A. Myllari, et al., ASP Conf. Ser. 316,
199 (2004).

4. T. Foster and B.Cooper, in TheDynamic Interstellar
Medium: A Celebration of the Canadian Galactic
Plane Survey, Ed. by R. Kothes, T. L. Landecker,
and A. G. Willis, ASP Conf. Ser. 438, 16 (2010);
arXiv:1009.3220 [astro-ph] (2010).

5. F. J. Kerr and D. Lynden-Bell,Mon. Not. R. Astron.
Soc. 221, 1023 (1986).

6. V. G. Surdin, Astron. Astrophys. Trans. 18, 367
(1999).

7. E. V. Glushkova, A. K. Dambis, A. M. Mel'nik,
and A. S. Rastorguev, Astron. Astrophys. 329, 514
(1998).

8. E.V. Glushkova, A. K. Dambis, and A. S. Rastorguev,
Astron. Astrophys. Trans. 18, 349 (1999).

9. B. Paczy\'nski and K. Z. Stanek, Astrophys. J. Lett.
494, L219 (1998); e-Print arXiv:astro-ph/9708080
(1997).

10. K. Z. Stanek, J. Kaluzny, A. Wysocka, and
I. Thompson, Acta Astron. 50, 191 (2000);
arXiv:astro-ph/9908041 (1999).

11. F. Eisenhauer, R. Sch\"odel, R. Genzel,
et al., Astrophys. J. Lett. 597, L121 (2003);
arXiv:astro-ph/0306220 (2003).

12. F. Eisenhauer, R. Genzel, T. Alexander, et al., Astrophys.
J. 628, 246 (2005); arXiv:astro-ph/0502129
(2005).

13. S. Gillessen, F. Eisenhauer, S. Trippe, et al., Astrophys.
J. Lett. 692, 1075 (2009); arXiv:0810.4674
[astro-ph] (2008).

14. S. Nishiyama, T. Nagata, S. Sato, et al., Astrophys.
J. 647, 1093 (2006); e-Print arXiv:astro-ph/0607408
(2006).

15. M. A. T. Groenewegen, A. Udalski, and G. Bono,
Astron. Astrophys. 481, 441 (2008), arXiv:0801.2652
[astro-ph] (2008).

16. S. Trippe, S. Gillessen, O. E. Gerhard, et al., Astron.
Astrophys. 492, 419 (2008); e-Print arXiv:0810.1040
[astro-ph] (2008).

17. S. Gillessen, F. Eisenhauer, T. K. Fritz, et al., Astrophys.
J. 707, L114 (2009); arXiv:0910.3069 [astro-
ph] (2009).

18. N. Matsunaga, T. Kawadu, S. Nishiyama, et al.,
Mon. Not. R. Astron. Soc. 399, 1709 (2009);
arXiv:0907.2761 [astro-ph] (2009).

19. M. Sato, M. J. Reid, A. Brunthaler, and
K. M. Menten, Astrophys. J. 720, 1055 (2010),
arXiv:1006.4218 [astro-ph] (2010).

20. I. S. Glass, P. A. Whitelock, R. M. Catchpole, and
M. W. Feast, Mon. Not. R. Astron. Soc. 273, 383
(1995).

21. A. C. Layden, R. B. Hanson, S. L. Hawley, et al.,
Astron. J. 112, 2110 (1996); arXiv:astro-ph/9608108
(1996).

22. R. Genzel, C. Pichon, A. Eckart, et al.,
Mon. Not. R. Astron. Soc. 317, 348 (2000);
arXiv:astro-ph/0001428 (2000).

23. M. A. T. Groenewegen and J. A. D. L. Blommaert,
Astron. Astrophys. 443, 143 (2005);
arXiv:astro-ph/0506338 (2005).

24. M. Shen and Z. Zhu, Chin. J. Astron. Astrophys. 7,
120 (2007).

25. A. M. Ghez, S. Salim, N. N. Weinberg, et al., Astrophys.
J. 689, 1044 (2008); arXiv:0808.2870 [astro-
ph] (2008).

26. D. J. Majaess, D. G. Turner, and D. J. Lane,
Mon. Not. R. Astron. Soc. 398, 263 (2009),
arXiv:0903.4206 [astro-ph] (2009).

27. M. J. Reid, K. M. Menten, X. W. Zheng, et al.,
Astrophys. J. 705, 1548 (2009); arXiv:0908.3637
[astro-ph] (2009).

28. E. Vanhollebeke, M. A. T. Groenewegen, and L. Girardi,
Astron. Astrophys. 498, 95 (2009).

29. M. R. Merrifield, Astron. J. 103, 1552 (1992).

30. C. R.Gwinn, J.M.Moran, andM. J. Reid, Astrophys.
J. 393, 149 (1992).

31. T. P. Gerasimenko, Astron. Rep. 48, 103 (2004).

32. J. M. Moran, M. J. Reid, and C. R. Gwinn, in
Astrophysical Masers, Ed. by A. W. Clegg and
G. E. Nedoluha, Lect. Notes Phys. 412, 244 (1993).

33. C. Babusiaux and G. Gilmore, Mon. Not. R. Astron.
Soc. 358, 1309 (2005); arXiv:astro-ph/0501383
(2005).

34. W. J. Maciel, Astrophys. Space Sci. 206, 285 (1993).

35. V. S. Avedisova, Astron. Rep. 49, 435 (2005).

36. F. Pont, M. Mayor, and G. Burki, Astron. Astrophys.
285, 415 (1994).

37. I. I. Nikiforov and I. V. Petrovskaya, Astron. Rep. 38,
642 (1994).

38. E. Bica, C. Bonatto, B. Barbuy, and S. Ortolani,
Astron. Astrophys. 450, 105 (2006); e-Print
arXiv:astro-ph/0511788 (2005).

39. A.S.Rastorguev,O. V. Durlevich, E.D. Pavlovskaya,
and A. A. Filippova, Astron. Lett. 20, 591 (1994).

40. A. K. Dambis, A. M. Mel'nik, and A. S. Rastorguev,
Astron. Lett. 21, 291 (1995).

41. V. V. Bobylev, A. T. Baikova, and S. V. Lebedeva,
Astron. Lett. 33, 571 (2007)]; arXiv:0709.4161 [astro-
ph] (2007).

42. B.W. Carney, J. P. Fulbright, D. M. Terndrup, et al.,
Astron. J. 110, 1674 (1995).

43. D. Huterer, D. D. Sasselov, and P. L. Schechter,
Astron. J. 110, 2705 (1995); arXiv:astro-ph/9508122
(1995).

44. M. Honma and Y. Sofue, Publ. Astron. Soc. Jpn. 48,
L103 (1996); arXiv:astro-ph/9611156 (1996).

45. M. W. Feast, Mon. Not. R. Astron. Soc. 284, 761
(1997).

46. M. Feast and P.Whitelock, Mon. Not. R. Astron. Soc.
291, 683 (1997); arXiv:astro-ph/9706293 (1997).

47. A. K. Dambis, Mon. Not. R. Astron. Soc. 396, 553
(2009).

48. M. R. Metzger, J. A. R. Caldwell, and P. L. Schechter,
Astron. J. 115, 635 (1998); arXiv:astro-ph/9710055
(1997).

49. C. Bonatto, E. Bica, B. Barbuy, and S. Ortolani,
in Globular Clusters-Guides to Galaxies, Ed. by
T. Richtler and S. Larsen (Springer, Berlin, Heidelberg,
2009), p. 209.

50. A. Udalski, Acta Astron. 48, 113 (1998);
arXiv:astro-ph/9805221 (1998).

51. M. J. Reid, K.M.Menten, X.W. Zheng, et al., Astrophys.
J. 700, 137 (2009); arXiv:0902.3913 [astro-ph]
(2009).

52. R. P. Olling and M. R. Merrifield, Mon. Not. R. Astron.
Soc. 297, 943 (1998); arXiv:astro-ph/9802034
(1998).

53. M. Feast, F. Pont, and P. Whitelock,Mon. Not. R. Astron.
Soc. 298, L43 (1998).

54. K. Z. Stanek and P.M.Garnavich, Astrophys. J. Lett.
503, L131 (1998), arXiv:astro-ph/9802121 (1998).

55. A. K. Dambis, in Variable Stars, the Galactic
Halo and Galaxy Formation, Ed. by C. Sterken,
N. Samus, and L. Szabados (Sternberg Astron.
Inst., Moscow Univ., Moscow, 2010), p. 177;
arXiv:1001.1428 [astro-ph] (2010).

56. D. H. McNamara, J. B. Madsen, J. Barnes, and
B. F. Ericksen, Publ. Astron. Soc. Pacif. 112, 202
(2000).

57. D. Majaess, Acta Astron. 60, 55 (2010);
arXiv:1002.2743 [astro-ph] (2010).

58. I. I. Nikiforov, in Small Galaxy Groups, Ed. by
M. J.Valtonen and C. Flynn, ASP Conf. Ser. 209, 403
(2000).

59. K. Ando, T. Nagayama, T. Omodaka, et al., Publ.
Astron. Soc. Jpn. 63, 45 (2011); arXiv:1012.5715
[astro-ph] (2012).

60. D. R. Alves, Astrophys. J. 539, 732 (2000);
arXiv:astro-ph/0003329 (2000).

61. A. Brunthaler, M. J. Reid, K. M. Menten, et al.,
Astron. Nachr. 332, 461 (2011); arXiv:1102.5350
[astro-ph] (2011).

62. M. J. Reid, in The Center of the Galaxy, Ed. by
M.Morris, p. 37 (1989).

63. Z. Malkin, J. Geodyn. 82, 325 (2008);
arXiv:physics/0703035 (2007).

64. Z. M. Malkin, Kinem. Phys. Celest. Bodies 27, 42
(2011); arXiv:1105.3837 [astro-ph] (2011).

\bigskip\flushright
Translated by N. Lipunova

\end{document}